\documentclass[useAMS,usegraphicx,usenatbib,usedcolumn,preprint]{aastex}
\usepackage{spr-astr-addons}
\usepackage{lscape}
\usepackage{latexsym}
\usepackage{graphics}
\usepackage{sidecap}
\usepackage{subfigure}
\usepackage{textcomp}
\usepackage{bbding}
\usepackage{epsf}
\usepackage{amsmath,amssymb}
\usepackage{sidecap}
\usepackage{multirow}
\usepackage{flafter}
\usepackage{url}
\usepackage{float}
\usepackage{color}
\usepackage{longtable}
\usepackage{rotating}
\usepackage{placeins}

\usepackage{epstopdf}
\makeatletter

\RequirePackage{color}

\newcommand{\NGC}{NGC\,55}

\newcommand{\HII}{\hbox{H\,{\sc ii}}}

\newcommand{\NUMBKG}{46}
\newcommand{\NUMHII}{11}
\newcommand{\NUMSNR}{6}

\begin{document}
\title{Radio-Continuum Study of the Nearby Sculptor Group Galaxies. Part~2: \NGC\ at $\lambda$=20, 13, 6 and 3~cm}

\shorttitle{\NGC\ at $\lambda$=20, 13, 6 and 3~cm}
\shortauthors{Andrew N. O'Brien et al.}

\author{Andrew N. O'Brien, Miroslav D. Filipovi\'c, Evan J. Crawford, Nicholas F. H. Tothill, Jordan D. Collier, Ain Y. De Horta, Graeme F. Wong, Danica Dra\v skovi\'c, Jeff L. Payne}
\affil{University of Western Sydney, Locked Bag 1797, Penrith, NSW 2751, Australia}
\author{Thomas G. Pannuti, Jared P. Napier, Stephen A. Griffith, Wayne D. Staggs}
\affil{Department of Earth and Space Sciences, Space Science Center, 235 Martindale Drive, Morehead State University, Morehead, KY 40351, USA}
\author{Srdjan Kotu\v s}
\affil{Department of Physics, University of Novi Sad, Trg Dositeja Obradovi\'ca 4, 21000 Novi Sad, Serbia}

\begin{abstract}
A series of new radio-continuum ($\lambda$=20, 13, 6 and 3~cm) mosaic images focused on the \NGC\ galactic system were produced using archived observational data from the Australia Telescope Compact Array. These new images are both very sensitive (down to rms=33~$\mu$Jy) and feature high angular resolution (down to $<$4\arcsec). Using these newly created images, 66 previously unidentified discrete sources are identified. Of these sources, \NUMBKG{} were classified as background sources, \NUMHII{} as \HII{} regions and \NUMSNR{} as supernova remnant candidates. This relatively low number of SNR candidates detected coupled with the low number of large \HII{} regions is consistent with the estimated low star formation rate of the galaxy at $0.06M_{\sun}$ year$^{-1}$. Our spectral index map shows that the core of galaxy appears to have a shallow spectral index between $\alpha = -0.2$ and $-0.4$. This indicates that the core of the galaxy is a region of high thermal radiation output.
\end{abstract}

\section{Introduction}

At $\sim2.08$~Mpc (scale of $\sim10$~pc$/\arcsec$) away \citep{2009ApJS..183...67D}, \NGC{} is situated between our own Local Group of galaxies and the nearby Sculptor Group \citep{2003A&A...404...93K}. The proximity is an advantage as it allows for \NGC{} to be examined in great detail. Previous radio-continuum studies of \NGC{} \citep{1996ApJS..103...81C,1991AJ....101..447P} utilised the Karl G. Jansky Very Large Array (VLA) in compact array configurations as their primary instrument and thus suffer from low resolution. As a result, these studies did not provide source lists of objects within the \NGC{} field.

Until the next generation of radio telescopes such as the Australian Square Kilometre Array Pathfinder (ASKAP), Karoo Array Telescope (KAT \& MeerKAT) and the Square Kilometre Array (SKA) become operational, we are restricted to consolidating a selection of \NGC{} radio observations. Part 1 of this paper (\citet{2012Ap&SS.340..133G}, Paper 1 hereafter) published a new set of highly sensitive and high-resolution radio-continuum images of the NGC~300 field at $\lambda$=20~cm, created by combining data from the Australia Telescope Compact Array (ATCA) and the VLA (also see \citet{2004A&A...425..443P}). In this paper, we examine all available archived radio-continuum observations of \NGC{} conducted with the ATCA and the VLA at $\lambda$=20, 13, 6 and 3~cm ($\nu$=1.4, 2.3, 5.5, 9.0~GHz) with the intention of merging these observations to create a single radio-continuum image following a similar methodology as presented in Paper 1. By combining a large amount of existing data using the latest generation of computing power we can create new images that feature both high angular resolution and excellent sensitivity. The newly constructed images are analysed and the differences between each map of \NGC{} created at the observed wavelengths are discussed.

In \S2 we describe the observational data and reduction techniques. In \S3 we present our new maps, a brief discussion and source list is given in \S4, and \S5 is the conclusion.

\section{DATA AND DATA REDUCTION}
\subsection{Observational Data}

In order to create a high-resolution and sensitive radio-continuum image, 29 observations from the ATCA and VLA were considered. These observations were selected from the Australian Telescope Online Archive (ATOA) and the National Radio Astronomy Observatory (NRAO) Science Data Archive. The observations which were selected and considered are summarised in Tables~\ref{table:NGC55-20-summary-obs}, \ref{table:NGC55-13-summary-obs}, \ref{table:NGC55-6-summary-obs} and \ref{table:NGC55-3-summary-obs}.

All ATCA projects, excluding C295, were conducted in mosaic mode with multiple pointings observed. All VLA observations and ATCA project C295 consisted of single pointings of \NGC{}. All images are primary beam corrected.

\begin{table*}
\tiny
\center
\caption[Summary of VLA and ATCA observations of NGC~55 at $\lambda=$20 cm used.]{Summary of VLA and ATCA observations of NGC~55 at $\lambda=$ 20 cm used in this study. \label{table:NGC55-20-summary-obs}}
\begin{tabular}{cccccccccc}
\hline
Instrument & Project & Dates & Array & $\nu$ & $\Delta \nu$ & Primary & Secondary & Integration & Included\\
& Code & & & (MHz) & (MHz) & Calibrator & Calibrator & Time (h) &in Image\\  \hline
VLA & AJ0107:A & 23 Mar 1984 & BC & 1465, 1515 & 50, 50 & 0134+329 & 0008-421 & 0.68 & N\\
VLA & AC0101:A & 13 Jul 1984 & CD & 1465, 1515 & 50, 50 & 0134+329 & 0025-263 & 0.12 & N\\
VLA & AS0199:A & 29 Aug 1984 & D & 1465, 1515 & 50, 50 & 0134+329 & 0022-423 & 0.23 & N\\
VLA & AB0342:A & 03 Nov 1985 & CD & 1465, 1515 & 50, 50 & 0134+329 & 0023-263 & 0.15 & N\\
VLA & AB0646:A & 03 Jul 1992 & CD & 1465, 1515 & 50, 50 & 0134+329 & 0008-421 & 2.51 & N\\
VLA & AB0646:B & 05 Jul 1992 & CD & 1465, 1515 & 50, 50 & 0134+329 & 0008-421 & 2.93 & N\\
ATCA & C1757 & 18-19 Feb 2009 & EW352 & 1384 & 128 & 1934-638 & 0008-421 & 61.62 & Y\\
ATCA & C1757 & 14-21 Nov 2008 & EW367 & 1384 & 128 & 1934-638 & 0008-421 & 188.00 & Y\\
ATCA & C1757 & 31 May-01 Jun 2008 & EW352 & 1384 & 128 & 1934-638 & 0008-421 & 65.57 & Y\\
ATCA & C1612 & 02-09 Oct 2006 & H75 & 1384 & 128 & 0407-658 & 0008-421 & 184.10 & Y\\
ATCA & C1341 & 07-08 Oct 2005 & EW214 & 1384 & 128 & 1934-638 & 0008-421 & 62.95 & Y\\
ATCA & C1341 & 18 Jul 2005 & H75 & 1384 & 128 & 1934-638 & 0008-421 & 21.32 & Y\\
ATCA & C287 & 25-26 Oct 1995 & 1.5D & 1344 & 128 & 1934-638 & 0008-421 & 65.07 & Y\\
ATCA & C287 & 12-13 Jan 1995 & 375 & 1380 & 128 & 1934-638 & 0008-421 & 63.40 & Y\\
ATCA & C287 & 01-02 Apr 1994 & 375 & 1380 & 128 & 1934-638 & 0008-421 & 54.13 & Y\\
ATCA & C287 & 01-02 Aug 1993 & 750D & 1380 & 128 & 1934-638 & 0008-421 & 61.29 & Y\\
ATCA & C295 & 24 Jun 1994 & 6C & 1380 & 128 & 1934-638 & 0023-263 & 0.99 & N\\
\hline
\end{tabular}
\end{table*}

\begin{table*}
\tiny
\center
\caption[Summary of VLA and ATCA observations of NGC~55 at $\lambda=$13 cm used.]{Summary of VLA and ATCA observations of NGC~55 at $\lambda=$13 cm used in this study. \label{table:NGC55-13-summary-obs}}
\begin{tabular}{cccccccccc}
\hline
Instrument & Project & Dates & Array & $\nu$ & $\Delta \nu$ & Primary & Secondary & Integration & Included\\
& Code & & & (MHz) & (MHz) & Calibrator & Calibrator & Time (h) &in Image\\  \hline
ATCA & C287 & 25-26 Oct 1995 & 1.5D & 2378 & 128 & 1934-638 & 0008-421 & 65.07 & Y\\
ATCA & C287 & 12-13 Jan 1995 & 375 & 2378 & 128 & 1934-638 & 0008-421 & 63.40 & Y\\
ATCA & C287 & 01-02 Apr 1994 & 375 & 2378 & 128 & 1934-638 & 0008-421 & 54.13 & Y\\
ATCA & C287 & 01-02 Aug 1993 & 750D & 2378 & 128 & 1934-638 & 0008-421 & 61.29 & Y\\
\hline
\end{tabular}
\end{table*}

\begin{table*}
\tiny
\center
\caption[Summary of ATCA CABB observations of NGC~55 at $\lambda=$6 cm used.]{Summary of ATCA CABB observations of NGC~55 at $\lambda=$ 6 cm used in this study. \label{table:NGC55-6-summary-obs}}
\begin{tabular}{cccccccccc}
\hline
Instrument & Project & Dates & Array & $\nu$ & $\Delta \nu$ & Primary & Secondary & Integration & Included\\
& Code & & & (MHz) & (MHz) & Calibrator & Calibrator & Time (h) &in Image\\  \hline
ATCA & C2421 & 15 Mar 2011 & 1.5A & 5500 & 2048 & 1934-638 & 0010-401 & 27.28 & Y\\
ATCA & C2421 & 20 Feb 2011 & EW352 & 5500 & 2048 & 1934-638 & 0010-401 & 26.91 & Y\\
ATCA & C2421 & 09-10 Nov 2010 & 750A & 5500 & 2048 & 1934-638 & 0010-401 & 31.04 & Y\\
ATCA & C1974 & 28 Mar 2010 & H168 & 5600 & 2048 & 1934-638 & 0008-421 & 136.80 & Y\\
ATCA & C1757 & 09-10 May 2009 & H168 & 5500 & 2048 & 1934-638 & 0022-423 & 370.00 & Y\\
\hline
\end{tabular}
\end{table*}

\begin{table*}
\tiny
\center
\caption[Summary of ATCA CABB observations of NGC~55 at $\lambda=$3 cm used.]{Summary of ATCA CABB observations of NGC~55 at $\lambda=$ 3 cm used in this study. \label{table:NGC55-3-summary-obs}}
\begin{tabular}{cccccccccc}
\hline
Instrument & Project & Dates & Array & $\nu$ & $\Delta \nu$ & Primary & Secondary & Integration & Included\\
& Code & & & (MHz) & (MHz) & Calibrator & Calibrator & Time (h) &in Image\\  \hline
ATCA & C2421 & 15 Mar 2011 & 1.5A & 9000 & 2048 & 1934-638 & 0010-401 & 27.28 & Y\\
ATCA & C2421 & 20 Feb 2011 & EW352 & 9000 & 2048 & 1934-638 & 0010-401 & 26.91 & Y\\
ATCA & C2421 & 09-10 Nov 2010 & 750A & 9000 & 2048 & 1934-638 & 0010-401 & 31.04 & Y\\
%ATCA & C1974 & 28 Mar 2010 & H168 & 9000 & 2048\\
%ATCA & C1757 & 09-10 May 2009 & H168 & 9000 & 2048\\
\hline
\end{tabular}
\end{table*}

\subsection{Data Reduction and Image Creation}

To create the best possible \NGC\ mosaic images, we included data from the fixed-position 6th antenna for all ATCA observations, as the large gaps in the {\it uv}-plane that dominate compact array configurations were filled mainly with data from observations in other array configurations. The inclusion of these long baselines resulted in images of high resolution with good {\it uv}-plane coverage.

The \textsc{miriad} \citep{miriad}, \textsc{aips} \citep{aips} and \textsc{karma} \citep{karma} software packages were used for data reduction and analysis. Because of the large volume of data, the \textsc{miriad} package was compiled to run on a 16-processor high-performance computer system.

\textsc{miriad} was used for all data reduction and imaging, however pre-processing in \textsc{aips} was required for the VLA data. This process consisted of importing the data into \textsc{aips} using the task \textsc{fillm}, and then splitting the sources into separate datasets with \textsc{split}. Using the task \textsc{uvfix}, source coordinates were converted from the B1950 to the J2000 reference frame. The task \textsc{fittp} was then used to export each source to a \textsc{fits} file. The \textsc{miriad} task \textsc{fits} was then used to import these fits files and convert them to \textsc{miriad} files. For the ATCA data, the task \textsc{atlod} was used to convert the raw observation files into \textsc{miriad} files.

Typical calibration and flagging procedures were then carried out \citep{miriad}, including the use of the guided automatic flagging task \textsc{pgflag}. Using the task \textsc{invert} with a robust weighting scheme, images were created for each ATCA and VLA project separately. Each image was then cleaned using the task \textsc{mossdi}. The \textsc{mossdi} task is a SDI clean algorithm designed for mosaic images \citep{1984A&A...137..159S}. To convolve a clean model the task \textsc{restor} was then used on each of the cleaned maps.  These images were created and visually inspected to assess the data quality.

Once the images had been verified to be free of errors, data observed at the same wavelength were combined in the {\it uv}-plane as mosaics by using the \textsc{invert} and providing the datasets of all observations as the input. This produced 4 dirty maps of all calibrated data of \NGC{} at 20, 13, 6 and 3~cm wavelengths, respectively. These dirty maps were then deconvolved using a more directed approach by providing \textsc{mossdi} with boxed regions around each visible source. This method produced superior images which were then restored to produce the combined images. The images were then cropped using \textsc{imsub} for analysis so that all images covered the same area of sky ($\sim 0.63 \arcdeg \times 0.41 \arcdeg$) centred at approximately 0h 15m 09s, $-39\arcdeg~12\arcmin~30\arcsec$ (J2000).

\section{RESULTS}
 
\begin{figure*} % example of high resolution/expanded arrayÉ maybe, find a better one
 	\center{\includegraphics[width=0.6\textwidth, angle=270]{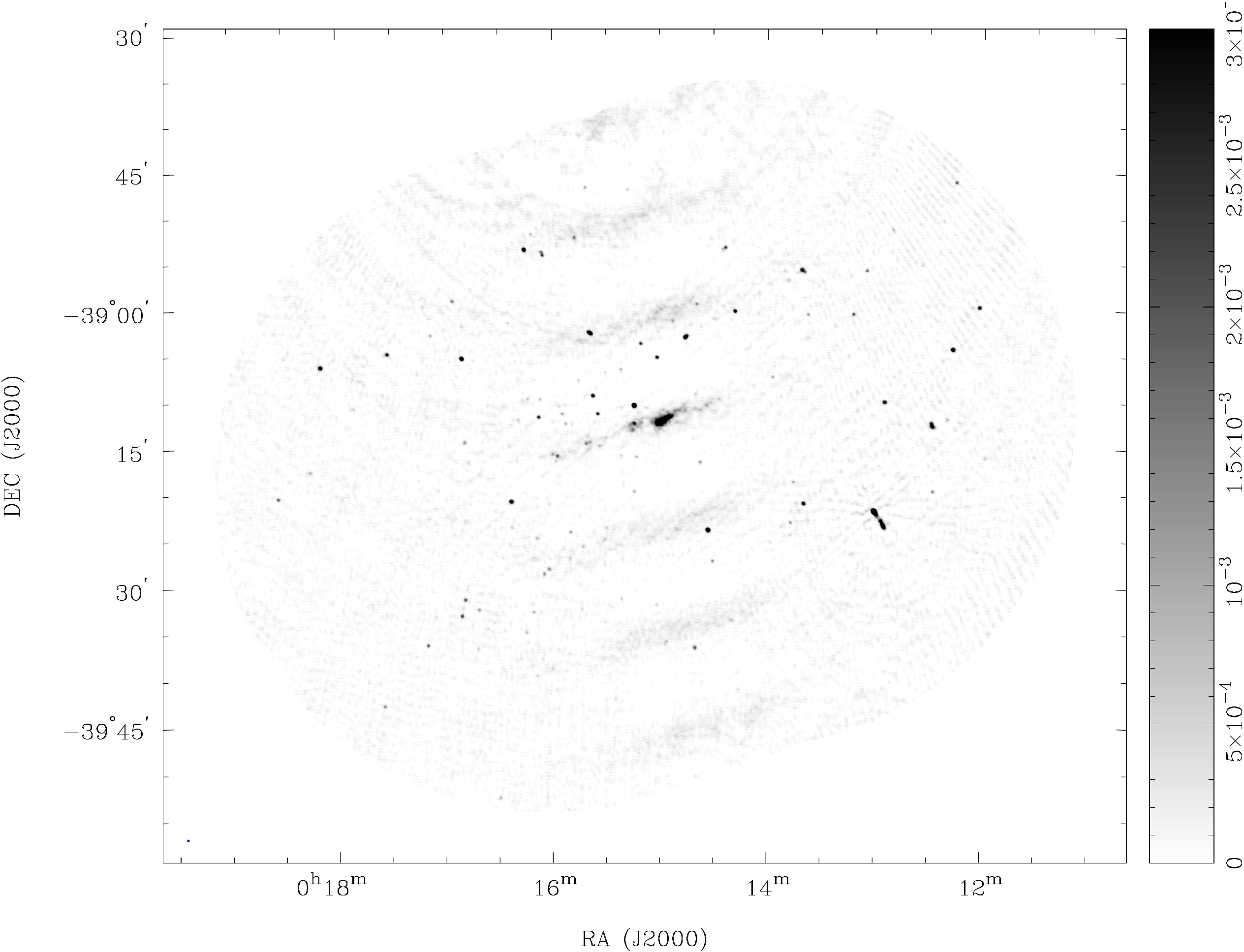}}
	\caption[Mosaic ATCA Project C287 total intensity image]{ATCA Project C287 radio-continuum mosaic of NGC~55 at $\lambda=20$~cm in Jy/beam. This image illustrates the high resolution produced by observations using a sparse array with long baselines. The synthesised beam is $7.93\arcsec \times 4.99\arcsec$ and the r.m.s. noise is 0.09~mJy/beam. \label{fig:C287}}
\end{figure*}

\begin{figure*} % example of low resolution/compact array
 	\center{\includegraphics[width=0.6\textwidth, angle=270]{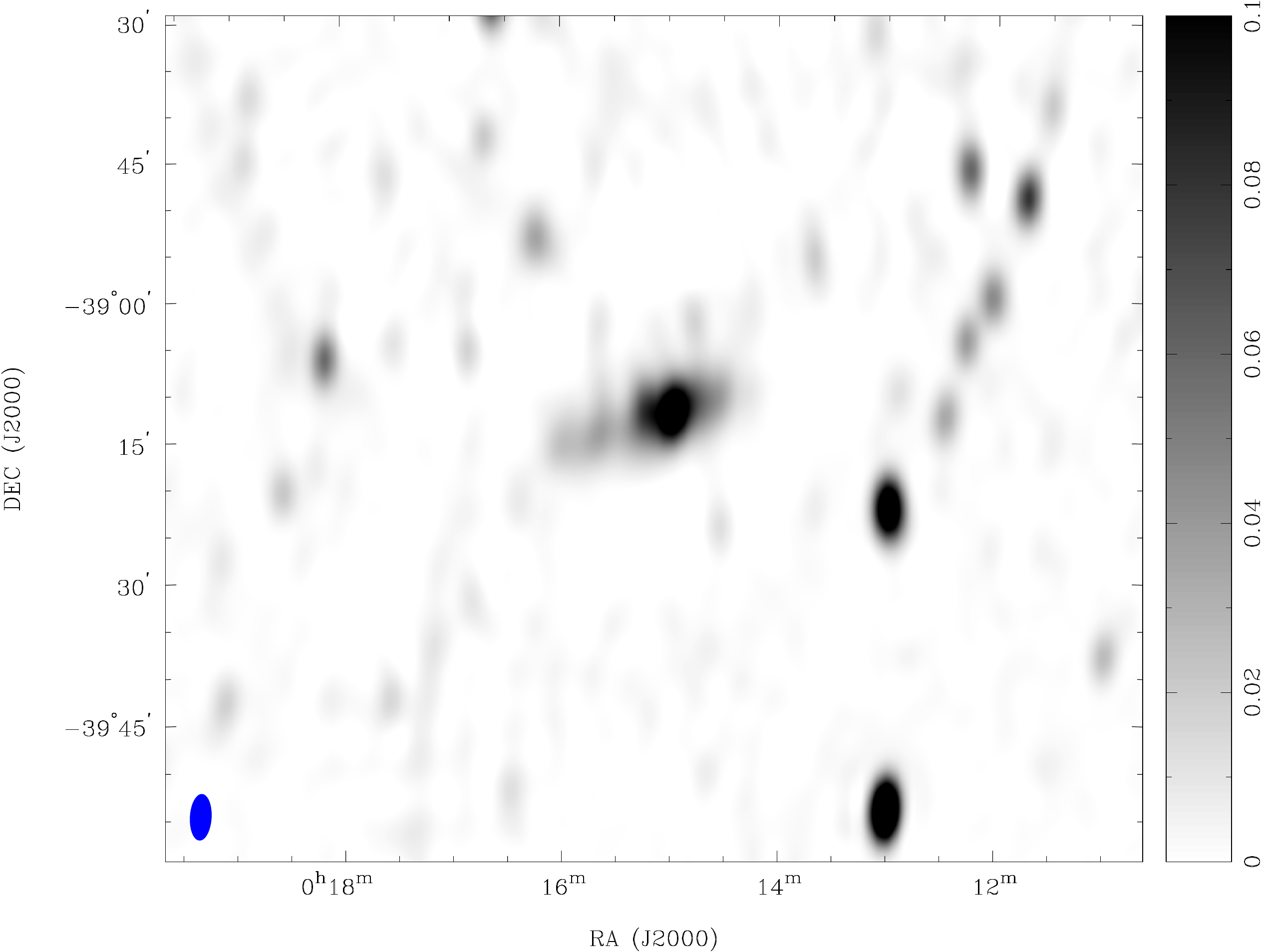}}
	\caption[Mosaic ATCA Project C1341 total intensity image]{ATCA Project C1341 radio-continuum mosaic of NGC~55 at $\lambda=20$~cm in Jy/beam. This image illustrates the low resolution produced by observations using a compact array with short baselines. The synthesised beam is $281.93\arcsec \times 188.82\arcsec$ and the r.m.s. noise is 2.13~mJy/beam. \label{fig:C1341}}
\end{figure*}

When visually inspecting the images produced from different observations, the effects of the different array configurations is apparent. As mentioned previously, compact arrays produce low resolution images with greater {\it uv}-coverage whereas larger arrays produce higher resolution images with less {\it uv}-coverage. While point sources are resolved to be much larger at lower resolutions, the advantage of the lower resolution is that they display extended emission throughout the field. This extended emission is lost in the high resolution images. The difference between an image produced using data from a compact array and an image produced using a larger configuration is obvious as can be seen in Figs.~\ref{fig:C287} and \ref{fig:C1341}. Ideal images are those which have high resolution but also contain extended emission data, which is where the technique of combining observations is advantageous.

After the initial inspection, it was determined that the data from the ATCA project C295 contained very poor {\it uv}-coverage and very limited hour angle coverage (0.99 hours). As a consequence, the final image contained several errors. Normally, these errors would disappear or lessen significantly when combined with other data observed in varying array configurations, as the majority of the gaps in the {\it uv}-coverage would be filled by the combined data. Since C295 was observed in one of the largest configurations (6C) and was the only observation at this configuration, there was insufficient data to fill the missing {\it uv}-coverage. Therefore, the data from C295 was excluded from the final image. Similarly, projects C1974 and C1757 were not included in the final 6~cm image because of their low resolution. As all of the 6~cm observations were conducted with the upgraded ATCA Compact Array Broadband Backend (CABB) and, at present, there exists no wide-bandwidth deconvolution algorithm for mosaics in \textsc{miriad}, we had to merge the images in the image plane using \textsc{immerge}. We found that when using this method, the low resolution data appeared to be weighted incorrectly and much of the high resolution data was overwhelmed by the low resolution data. Thus, these data were not used to create the final 6~cm image.

\begin{figure*}
 	\center{\includegraphics[width=0.5\textwidth, angle=270]{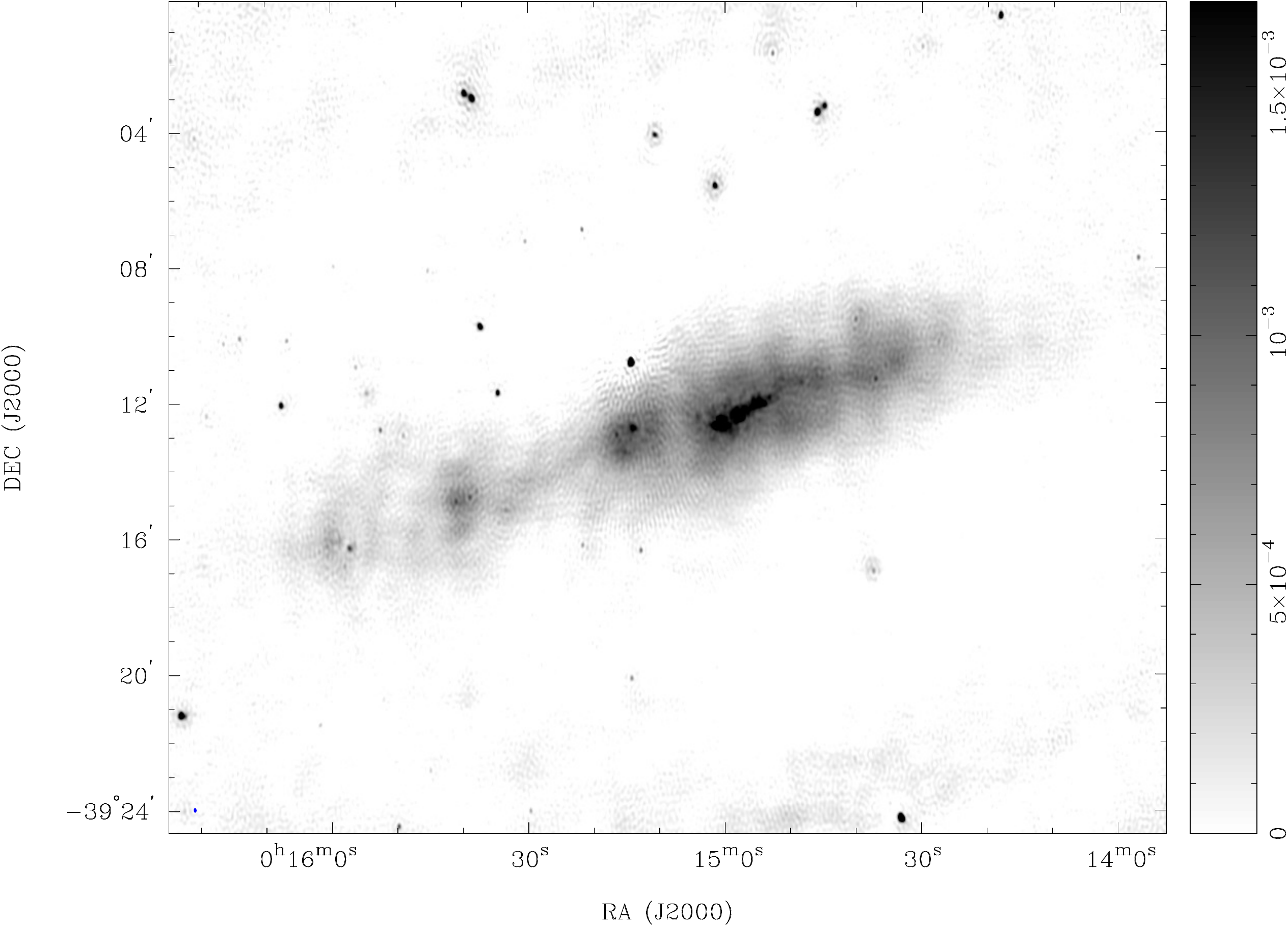}}
	\caption[Cropped total intensity image of all ATCA data]{Cropped total intensity image of all ATCA data (excluding C295) of NGC~55 at $\lambda=20$~cm in Jy/beam. The synthesised beam is $7.94\arcsec \times 4.99\arcsec$ and the r.m.s. noise is 0.12~mJy/beam. \label{fig:20cm-imsub}}
\end{figure*}

In addition, the VLA observations were also excluded from the final 20~cm image (Fig.~\ref{fig:20cm-imsub}) as the inclusion of this data overpowered the extended emission from the ATCA observations. Excluding this data was of little consequence as the main advantage of VLA data is the higher resolution, however the largest VLA array configuration used to observe NGC~55 was a single observation in the the BC array which has a maximum baseline separation of approximately 7.6~km. This is relatively close to that of the 6~km maximum baseline at ATCA and so we decided the preservation of the extended emission data was worth more than the slight increase in resolution that would be gained by including the VLA data. Information on the final images used in this paper is given in Table~\ref{table:image-details}.

%%%%%%%%%%%% Table 5
\begin{table*}
\center
\caption[Image details of ATCA projects of NGC~55 at 20, 13, 6 and 3~cm]{Image details of ATCA single and merged projects of NGC~55 mosaics at 20, 13, 6 and 3~cm that were used to produce the merged images used for the measurements given in this paper. Merged images were created using joint-deconvolution of all projects.\label{table:image-details}}
\begin{tabular}{lcccccc}
\hline
ATCA & Centre $\nu$ & Synthesised Beam & r.m.s & Figure\\
Project & (MHz) & FWHM ($\arcsec$) & (mJy/beam)\\
\hline
C287 & 1362 & $7.93 \times 4.99$ & 0.09 & \ref{fig:C287} \\
%C295 & 1380 & $29.75 \times 5.74$ & 0.20 & \ref{fig:C295} \\
C1341 & 1384 & $281.93 \times 188.82$ & 2.13 & \ref{fig:C1341} \\
C1612 & 1384 & $426.14 \times 279.93$ & 1.08 & -- \\
C1757 & 1384 & $29.29 \times 24.04$ & 0.99 & -- \\
%AB0342:A & 1490 & $62.90 \times 37.24$ & 2.91 & \ref{fig:AB0342_a} \\
%AB0646:A & 1490 & $57.12 \times32.61$ & 1.65 & \ref{fig:AB0646_a} \\
%AB0646:B & 1490 & $53.97 \times 33.41$ & 1.34 & \ref{fig:AB0646_b} \\
%AC0101:A & 1490 & $66.30 \times 37.17$ & 2.04 & \ref{fig:AC0101_a} \\
%AC0308:XX & 1400 & $72.45 \times 45.50$ & 8.13 & \ref{fig:AC0308_xx} \\
%AJ0107:A & 1490 & $20.04 \times 11.30$ & 1.03 & \ref{fig:AJ0107_a} \\
%AS0199:A & 1490 & $141.7 \times 36.68$ & 1.85 & \ref{fig:AS0199_a} \\
Merged 20~cm & 1366 & $7.94 \times 4.99$ & 0.09 & \ref{fig:20cm-imsub} \\
%All VLA 20~cm & 1430 & $18.18 \times 10.09$ & 0.14 & \ref{fig:all_VLA} \\
%All 20~cm & 1398 & $7.94 \times 4.99$  & 0.12 & \ref{fig:all_20cm} \\
\hline
C287 & 2378 & $18.29 \times 16.96$ & 0.18 & \ref{fig:13cm-imsub} \\
\hline
C2421 & 5500 & $5.43 \times 4.68$ & 0.03 & \ref{fig:6cm-imsub} \\
%C1974 & 5500 & $43.74 \times 30.15$ & 0.18 & \ref{fig:C1974_6cm} \\
%C1757 & 5500 & $45.31 \times 29.86$ & 0.07 & \ref{fig:C1757_6cm} \\
%Merged 6~cm & 5500 & $5.43 \times 4.68$ & 0.03 & \ref{fig:6cm-final} \\
\hline
C2421 & 9000 & $5.26 \times 3.62$ & 0.05 & \ref{fig:3cm-imsub} \\
\hline
\end{tabular}
\end{table*}
%%%%%%%%%%%%%%%%%%%%%%%%%%%%%%%%%%%%%%%%%

%%%%%%%%%%%%%%%% FINALS

\begin{figure*}
 	\center{\includegraphics[width=0.5\textwidth, angle=270]{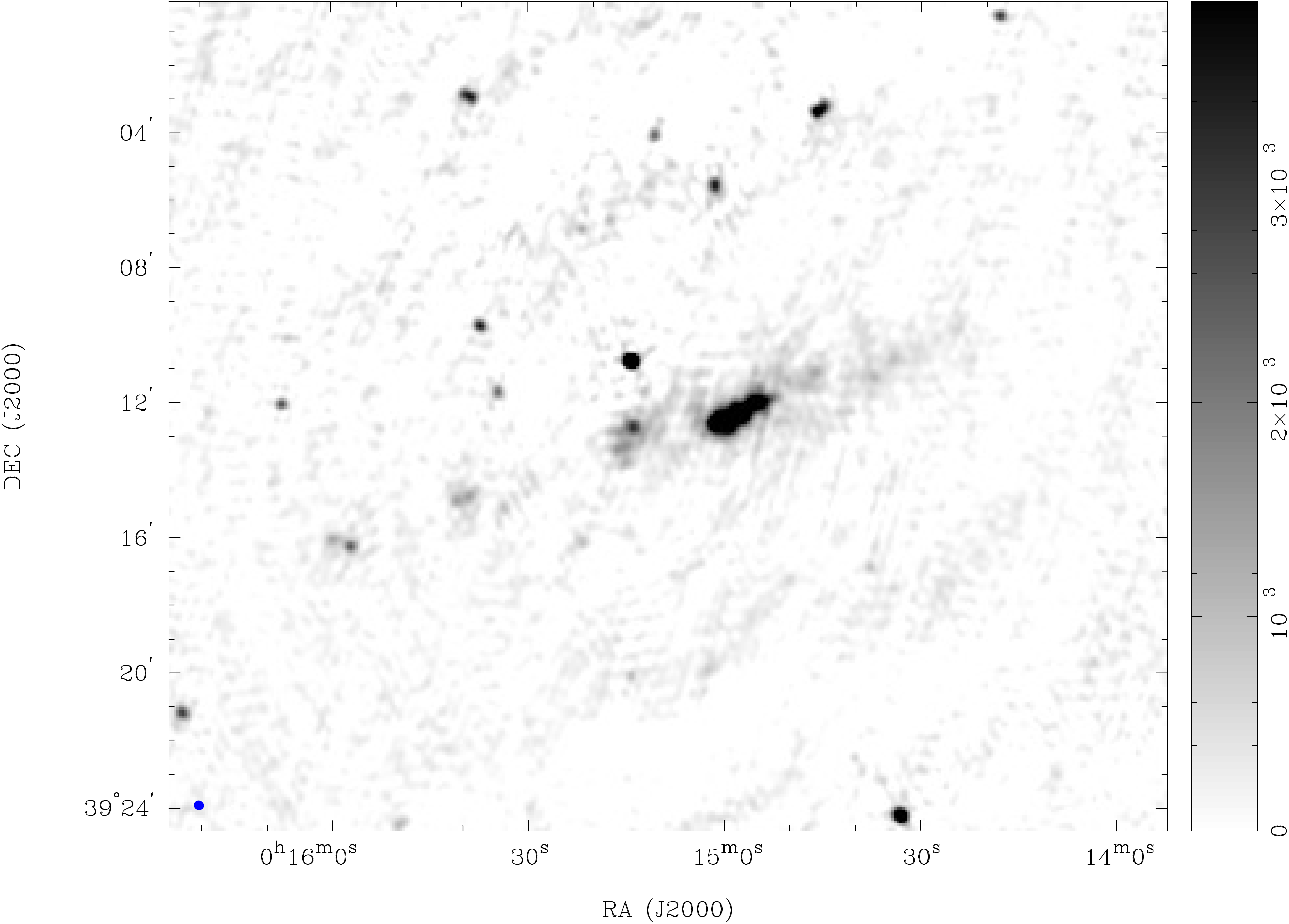}}
	\caption[Cropped total intensity image of all ATCA data]{Cropped total intensity image of all ATCA data of NGC~55 at $\lambda=13$~cm in Jy/beam. The synthesised beam is $18.29\arcsec \times 16.96\arcsec$ and the r.m.s. noise is 0.09~mJy/beam. \label{fig:13cm-imsub}}
\end{figure*}

\begin{figure*}
 	\center{\includegraphics[width=0.5\textwidth, angle=270]{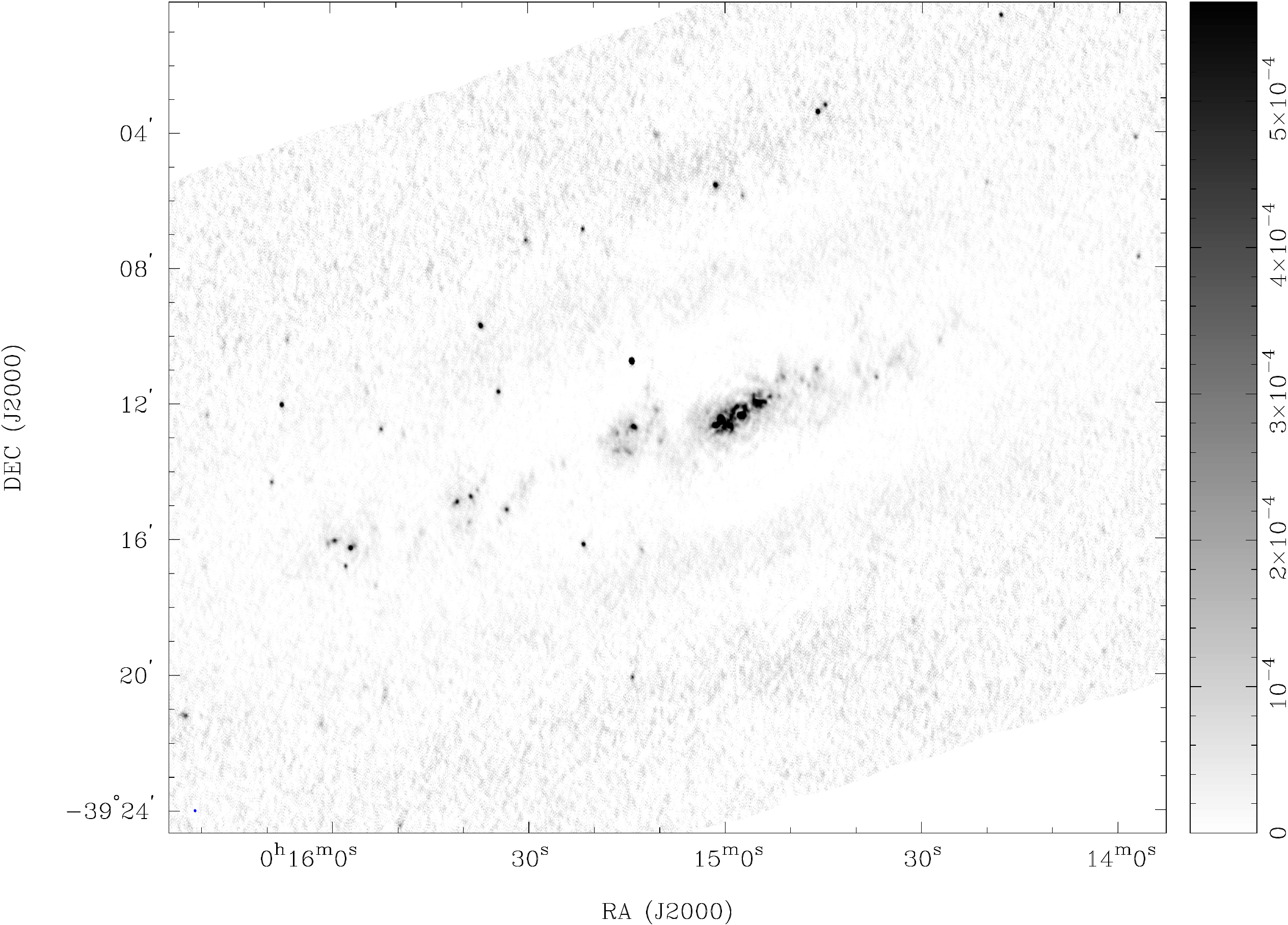}}
	\caption[Cropped total intensity image of all ATCA data]{Cropped total intensity image of all ATCA data of NGC~55 at $\lambda=6$~cm in Jy/beam. The synthesised beam is $5.43\arcsec \times 4.68\arcsec$ and the r.m.s. noise is 0.03~mJy/beam. \label{fig:6cm-imsub}}
\end{figure*}

\begin{figure*}
 	\center{\includegraphics[width=0.5\textwidth, angle=270]{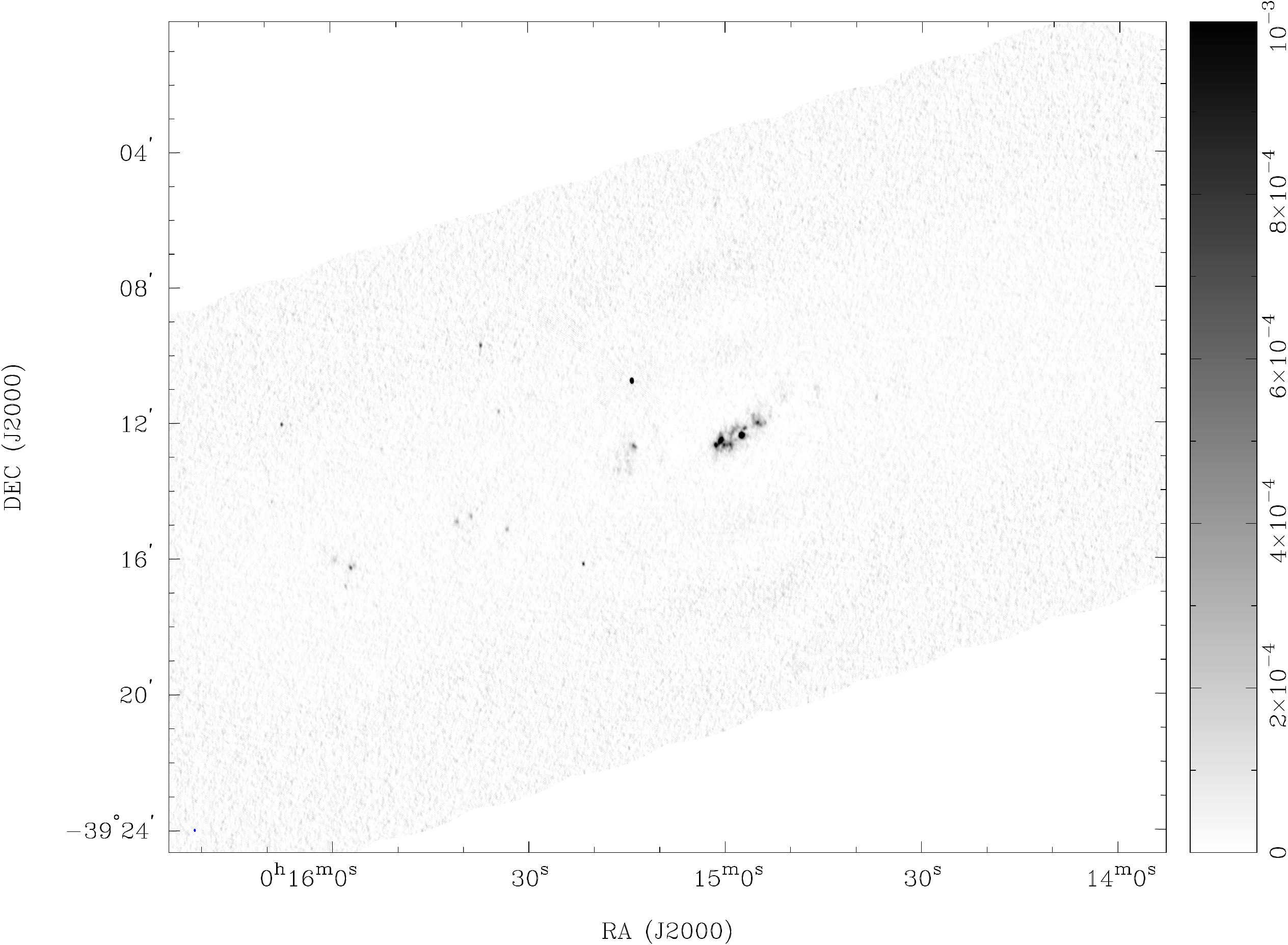}}
	\caption[Cropped total intensity image of all ATCA data]{Cropped total intensity image of all ATCA data of NGC~55 at $\lambda=3$~cm in Jy/beam. The synthesised beam is $5.26\arcsec \times 3.62\arcsec$ and the r.m.s. noise is 0.05~mJy/beam. \label{fig:3cm-imsub}}
\end{figure*}

\section{DISCUSSION}

\subsection{Discrete Sources Within The Field Of \NGC}

A total of 66 sources above $5\sigma$ (0.45, 0.92, 0.15, 0.25~mJy/beam at wavelengths of 20, 13, 6 and 3~cm, respectively) were identified within the field of NGC~55. These sources were catalogued by position and flux density. 33 of these sources were detected at more than one of the observed wavelengths and consequently their spectral index, $\alpha$, was estimated using the equation $S\propto\nu^{\alpha}$. Simple linear regression was then performed to determine the best-fit value for $\alpha$ between the measured frequencies. This catalogue was then compared to optical \citep{trove.nla.gov.au/work/9368097}, infrared \citep{2009ApJ...703..517D, 2003AJ....125..525J} and X-ray images \citep{2013AAS...22124921P, 2006MNRAS.370...25S} of the \NGC{} field. Source coincidences were accepted for sources within 1 beam of the highest resolution radio image being compared.

Based on their location and comparison with the images at other frequency bands, the sources were classified as either background (BKG) sources, supernova remnants (SNR), or ionised hydrogen regions (\HII). Sources outside of the extended emission boundary seen in Fig.~\ref{fig:20cm-imsub} were classified as background sources. Sources inside the boundary were classified as intrinsic (INTR) sources which are further classified as either SNRs or \HII{} regions. If X-ray radiation was detected from the intrinsic source, then it was classified as a SNR candidate since the violent explosions and shockwaves produced by a supernova are known to emit strong X-rays. Also, SNRs usually exhibit steeper spectral index of about $\alpha < -0.5$ \citep{1998A&AS..130..421F}. Sources emitting infrared radiation but no detectable X-rays were classified as \HII\ regions. Of the 66 detected radio-continuum sources, \NUMBKG{} were classified as background sources, \NUMHII{} as \HII\ regions, and \NUMSNR{} as SNR candidates. The full list of sources with positions, flux densities, spectral index and classifications can be found in Table~\ref{table:point-source-list}.

The low number of detected SNR candidates suggests that the star formation rate for \NGC{} is relatively low. To estimate the high mass star formation rate, we used the relation described by \citet{1983ApJ...272...54K}. We adjusted the estimated H$\alpha$ luminosity of \NGC{} given by \citet{1996AJ....112.1429H} to $4.08\times10^{40}$ergs s$^{-1}$ to reflect our adopted distance of 2.08 Mpc. The relation provided estimated high mass star formation rate of $0.06 M_{\sun}$ year$^{-1}$. Compared to the star formation rates of other galaxies in the Sculptor Group (the highest being NGC 253 at $0.20 M_{\sun}$ year$^{-1}$, and the lowest being NGC 45 at $0.01 M_{\sun}$ year$^{-1}$), the star formation rate of \NGC{} is comparatively low.

\begin{figure}[h]
	\center{\includegraphics[clip=true,width=\columnwidth,trim=2cm 0cm 2cm 0cm]{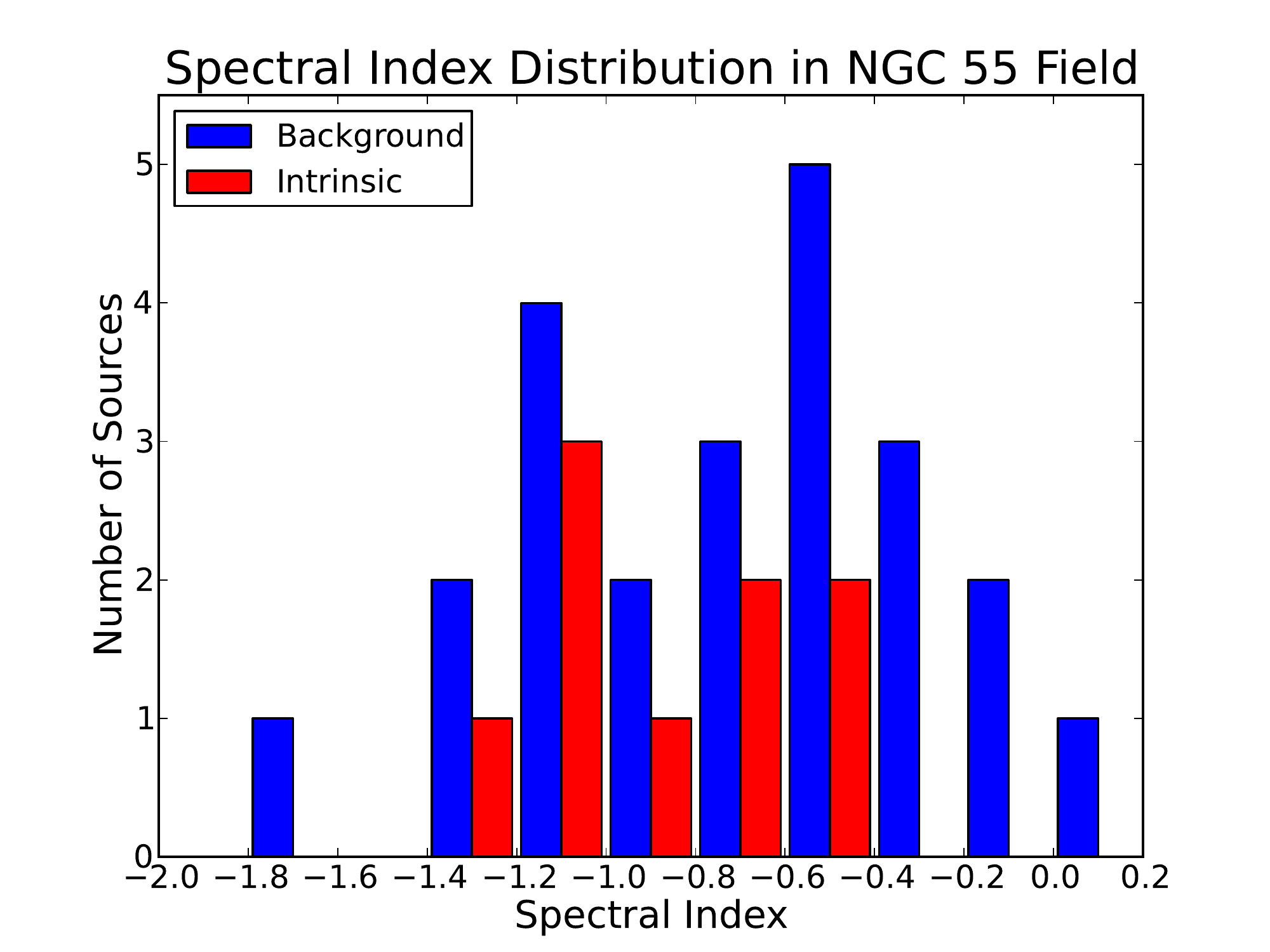}}
	\caption[Spectral index distribution of sources]{Histogram of the spectral index distribution of sources in the NGC~55 field with bin widths of 0.2. Sources with high spectral index uncertainties ($>\pm0.4$) were excluded. \label{fig:histogram}}
\end{figure}

In Fig.~\ref{fig:histogram} we show the spectral index distribution of all radio-continuum sources found in this study (Table~\ref{table:point-source-list}; Col.~9). Here, we note that the spectral index alone cannot successfully distinguish between various type of sources. Of the 66 sources reported here, 10 have an estimated spectral index of steeper than $-0.85$, classifying them as candidates for compact steep spectrum (CSS) sources (Table~\ref{table:point-source-list}; Col.~13).

\begin{table*}
\tiny
\center
\caption{List of point sources in the NGC~55 field at $\lambda=$20, 13, 6 and 3~cm. RA (3) and Dec (4) are in J2000 coordinates. Column 9 is the best fit spectral index for all flux measurements of a source. Blank cells indicate no measurement was detected. Asterisks (*) indicate no coverage was available. Daggers ($^\dagger$) indicate sources that were resolved as a single source at $\lambda=13$~cm but as multiple sources at other wavelengths. These sources were therefore not included in the estimation of $\alpha$. \label{table:point-source-list}}
\begin{tabular}{ccccccccccccl}
\hline
\hline
1 & 2 & 3 & 4 & 5 & 6 & 7 & 8 & 9 & 10 & 11 & 12 & 13\\
Index & Source & RA & Dec & S$_{20cm}$ & S$_{13cm}$ & S$_{6cm}$ & S$_{3cm}$ & $\alpha$ & Optical & IR & X-Ray & Source\\
& Name & (h \space m \space s) & (\textdegree \space \arcmin \space \arcsec) & (mJy) & (mJy) & (mJy) & (mJy) & & ID & ID & ID & Class\\ 
 \hline
1 & J001357-390742 & 00:13:57 & -39:07:42 & 1.05 & & 0.40 & & $-0.69\pm0.14$ & * & Y & & BKG \\
2 & J001358-390409 & 00:13:58 & -39:04:09 & & & 0.41 & 0.31 & $-0.55\pm0.40$ & * & & & BKG \\
3 & J001418-390034 & 00:14:18 & -39:00:34 & 5.47 & 3.07 & 0.89 & & $-1.32\pm0.12$ & * & Y & & BKG (CSS) \\
4 & J001420-390531 & 00:14:20 & -39:05:31 & & & 0.21 & & & & & & BKG \\
\smallskip
5 & J001427-391012 & 00:14:27 & -39:10:12 & & & 0.15 & & & Y & Y & & \HII{} \\
6 & J001430-390129 & 00:14:30 & -39:01:29 & 0.50 & & & & & * & Y & & BKG \\
7 & J001431-391829 & 00:14:31 & -39:18:29 & & & 0.19 & & & Y & & & BKG \\
8 & J001433-392415 & 00:14:33 & -39:24:15 & 11.84 & 10.77 & & & $-0.17\pm0.36$ & * & & & BKG \\
9 & J001437-391118 & 00:14:37 & -39:11:18 & 1.89 & 1.25 & 0.37 & 0.25 & $-1.13\pm0.10$ & Y & & Y & SNR\\
\smallskip
10 & J001437-391659 & 00:14:37 & -39:16:59 & 0.81 & & & & & & & & BKG \\
11 & J001440-390934 & 00:14:40 & -39:09:34 & 1.15 & & & & & Y & & & \HII{}\\
12 & J001445-390315 & 00:14:45 & -39:03:15 & 2.96 & 6.77$^\dagger$ & 0.73 & & $-1.01\pm0.14$ & * & & Y & BKG (CSS) \\
13 & J001446-390327 & 00:14:46 & -39:03:27 & 6.16 & 6.77$^\dagger$ & 1.82 & & $-0.88\pm0.14$ & * & & & BKG (CSS)\\
14 & J001446-391104 & 00:14:46 & -39:11:04 & & & 0.43 & & & Y & Y & & \HII{}\\
\smallskip
15 & J001447-391114 & 00:14:47 & -39:11:14 & & 1.96 & & & & Y & & Y & SNR\\
16 & J001447-391133 & 00:14:47 & -39:11:33 & & & 0.20 & & & Y & Y & & \HII{}\\
17 & J001448-391123 & 00:14:48 & -39:11:23 & & & 0.25 & & & Y & & & \HII{}\\
18 & J001451-391118 & 00:14:51 & -39:11:18 & & & 0.42 & & & Y & & & \HII{}\\
19 & J001452-391152 & 00:14:52 & -39:11:52 & & & 0.32 & & & Y & Y & & \HII{}\\
\smallskip
20 & J001453-390142 & 00:14:53 & -39:01:42 & 0.97 & & & & & * & Y & & BKG \\
21 & J001457-390556 & 00:14:57 & -39:05:56 & & & 0.33 & & & & & & BKG \\
22 & J001457-390559 & 00:14:57 & -39:05:59 & & & 0.18 & & & & & & BKG \\
23 & J001502-390537 & 00:15:02 & -39:05:37 & 3.65 & 4.37 & 1.88 & & $-0.52\pm0.36$ & Y & Y & Y & BKG \\
24 & J001511-390403 & 00:15:11 & -39:04:03 & 2.19 & 2.46 & 0.18 & & $-1.92\pm0.92$ & * & & & BKG (CSS)\\
\smallskip
25 & J001513-391624 & 00:15:13 & -39:16:24 & 1.10 & & 0.24 & & $-1.09\pm0.14$ & & Y & & BKG (CSS)\\
26 & J001514-391049 & 00:15:14 & -39:10:49 & 23.31 & 20.26 & 13.68 & 8.80 & $-0.51\pm0.08$ & & & & BKG \\
27 & J001514-391246 & 00:15:14 & -39:12:46 & 3.51 & 4.67 & 1.62 & 0.89 & $-0.81\pm0.27$ & Y & Y & Y & SNR\\
28 & J001514-392009 & 00:15:14 & -39:20:09 & 0.88 & & 0.54 & & $-0.35\pm0.14$ & * & & & BKG \\
29 & J001516-391308 & 00:15:16 & -39:13:08 & 1.38 & & & & & Y & Y & & \HII{}\\
\smallskip
30 & J001518-390736 & 00:15:18 & -39:07:36 & & & 0.12 & & & Y & & & BKG \\
31 & J001522-390655 & 00:15:22 & -39:06:55 & 0.98 & 1.25 & 0.72 & & $-0.26\pm0.30$ & & & & BKG \\
32 & J001522-391613 & 00:15:22 & -39:16:13 & 0.81 & 1.05 & 1.14 & 0.97 & $0.10\pm0.10$ & & & Y & BKG \\
33 & J001530-390715 & 00:15:30 & -39:07:15 & 0.58 & & 0.58 & & $-0.01\pm0.14$ & Y & Y & Y & BKG \\
34 & J001530-392403 & 00:15:30 & -39:24:03 & 0.63 & & & & & * & & & BKG \\
\smallskip
35 & J001533-391511 & 00:15:33 & -39:15:11 & 1.00 & & 0.67 & 0.42 & $-0.42\pm0.13$ & * & Y & Y & SNR\\
36 & J001535-391143 & 00:15:35 & -39:11:43 & 3.05 & 2.38 & 1.02 & 0.42 & $-1.04\pm0.17$ & * & Y & & BKG (CSS)\\
37 & J001537-390946 & 00:15:37 & -39:09:46 & 4.53 & 3.88 & 1.94 & 0.87 & $-0.86\pm0.17$ & * & Y & Y & BKG (CSS)\\
38 & J001538-390301 & 00:15:38 & -39:03:01 & 6.32 & & & & & * & & & BKG \\
39 & J001538-391437 & 00:15:38 & -39:14:37 & 1.08 & & 0.22 & & $-1.13\pm0.14$ & * & & & INTR\\
\smallskip
40 & J001539-390258 & 00:15:39 & -39:02:58 & & 4.67 & & & & * & & & BKG \\
41 & J001539-391448 & 00:15:39 & -39:14:48 & 1.69 & 2.16$^\dagger$ & 0.74 & 0.43 & $-0.69\pm0.10$ & * & Y & & INTR\\
42 & J001539-391533 & 00:15:39 & -39:15:33 & & & 0.24 & & & * & Y & & \HII{}\\
43 & J001540-390252 & 00:15:40 & -39:02:52 & 4.97 & & & & & * & Y & & BKG \\
44 & J001540-391452 & 00:15:40 & -39:14:52 & & 2.16 & & & & * & Y & & \HII{}\\
\smallskip
45 & J001541-391457 & 00:15:41 & -39:14:57 & 2.03 & 2.16$^\dagger$ & 0.84 & 0.46 & $-0.76\pm0.12$ & * & Y & Y & SNR\\
46 & J001545-390807 & 00:15:45 & -39:08:07 & 0.44 & & & & & * & Y & & BKG \\
47 & J001549-391259 & 00:15:49 & -39:12:59 & 0.59 & & & & & * & Y & & BKG \\
48 & J001550-392429 & 00:15:50 & -39:24:29 & 1.32 & 1.21 & 0.32 & & $-1.07\pm0.39$ & * & & & BKG (CSS)\\
49 & J001552-391248 & 00:15:52 & -39:12:48 & 1.00 & & 0.51 & & $-0.48\pm0.14$ & * & Y & Y & BKG \\
\smallskip
50 & J001552-392029 & 00:15:52 & -39:20:29 & & & 0.18 & & & * & & & BKG \\
51 & J001552-392042 & 00:15:52 & -39:20:42 & & & 0.19 & & & * & & & BKG \\
52 & J001555-391143 & 00:15:55 & -39:11:43 & 0.67 & & & & & * & & & BKG \\
53 & J001556-391057 & 00:15:56 & -39:10:57 & 0.50 & & & & & * & & & BKG \\
54 & J001557-391618 & 00:15:57 & -39:16:18 & 1.76 & 2.86 & 1.43 & 0.83 & $-0.46\pm0.28$ & * & Y & & \HII{}\\
\smallskip
55 & J001558-391650 & 00:15:58 & -39:16:50 & & & 0.55 & 0.33 & $-1.02\pm0.40$ & * & Y & & INTR\\
56 & J001600-390758 & 00:16:00 & -39:07:58 & 0.23 & & & & & * & Y & & BKG \\
57 & J001600-391605 & 00:16:00 & -39:16:05 & & 1.58 & 0.74 & 0.27 & $-1.29\pm0.30$ & * & Y & Y & SNR\\
58 & J001602-392128 & 00:16:02 & -39:21:28 & 0.30 & & 0.22 & & $-0.22\pm0.14$ & * & Y & Y & BKG \\
59 & J001607-391009 & 00:16:07 & -39:10:09 & 0.66 & & 0.28 & & $-0.63\pm0.14$ & * & & Y& BKG \\
\smallskip
60 & J001608-391203 & 00:16:08 & -39:12:03 & 3.23 & 2.63 & 1.51 & 0.86 & $-0.69\pm0.10$ & * & & & BKG \\
61 & J001609-391420 & 00:16:09 & -39:14:20 & & & 0.46 & 0.25 & $-1.27\pm0.40$ & * & & & BKG (CSS)\\
62 & J001614-391005 & 00:16:14 & -39:10:05 & 0.73 & & & & & * & & & BKG \\
63 & J001616-391014 & 00:16:16 & -39:10:14 & 0.40 & & & & & * & & & BKG \\
64 & J001619-391221 & 00:16:19 & -39:12:21 & 0.43 & & 0.24 & & $-0.44\pm0.14$ & * & & & BKG \\
\smallskip
65 & J001622-392111 & 00:16:22 & -39:21:11 & 6.34 & 3.40 & 0.63 & & $-1.69\pm0.24$ & * & & & BKG (CSS)\\
66 & J001624-390151 & 00:16:24 & -39:01:51 & 0.55 & & & & & * & && BKG \\

\end{tabular}
\end{table*}

\subsection{Spectral Index Map}

\begin{figure*}
 	\center{\includegraphics[clip=true, trim=1.5cm 2cm 2.6cm 1.5cm, width=0.8\textwidth]{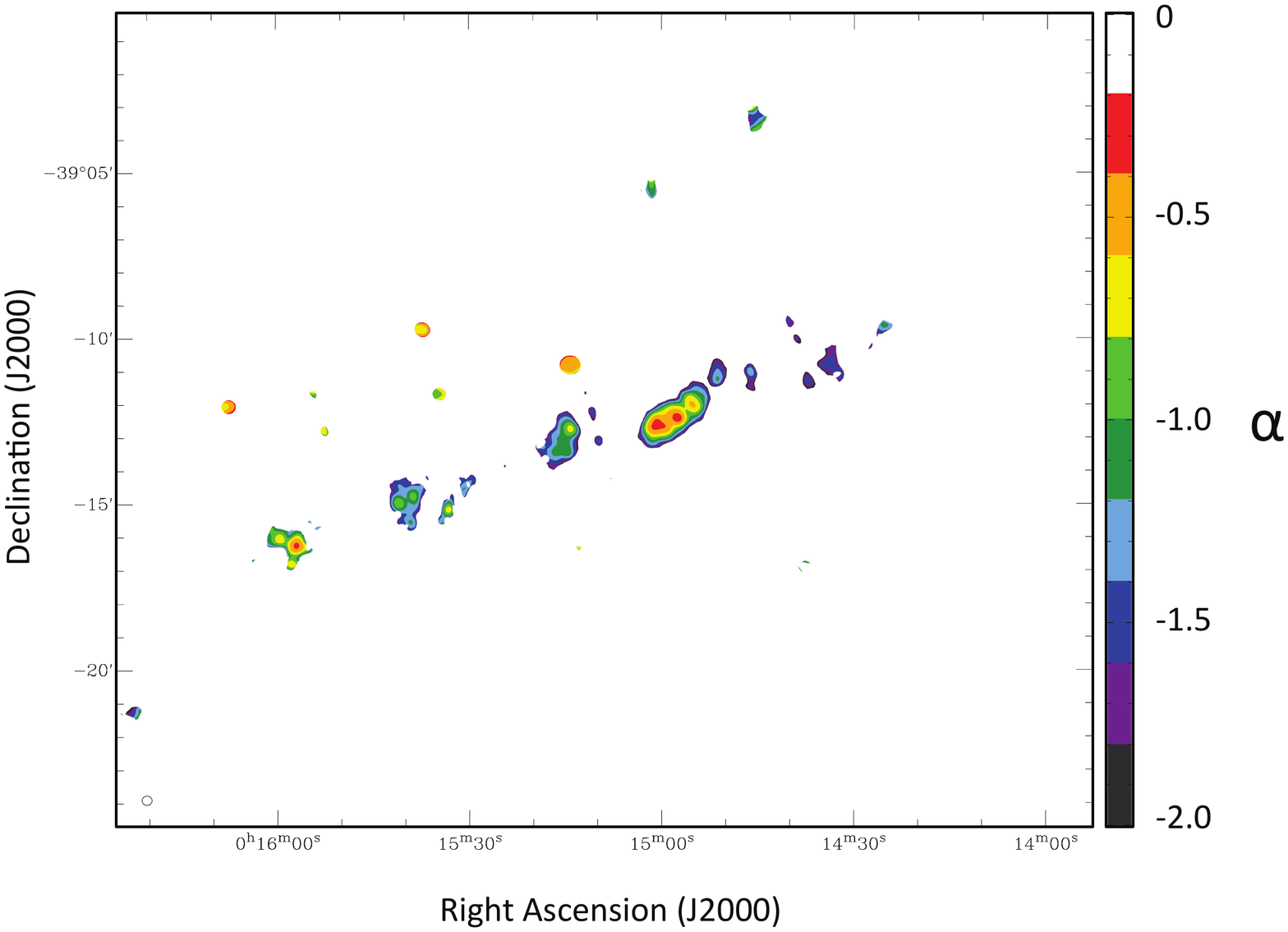}}
	\caption[Spectral index pixel map of NGC~55]{Spectral index pixel map of NGC~55. Image is in terms of spectral index $\alpha$. The colour bar on the right reflects the spectral index value multiplied by 1000. The synthesised beam is $18.29\arcsec \times 16.96\arcsec$. \label{fig:specindexmap}}
\end{figure*}

We show in Fig.~\ref{fig:specindexmap} an image where each pixel represents the spectral index calculated across all 4 observed frequencies. This image was created based on measurements from Figs. \ref{fig:20cm-imsub}, \ref{fig:13cm-imsub}, \ref{fig:6cm-imsub} and \ref{fig:3cm-imsub} after convolving the images to the largest beam size. Pixels below the noise level are ignored.

The core of galaxy is shown to have a shallow spectral index of between $-0.2$ and $-0.4$ (shown as red pixels). This indicates that the core of the galaxy is a region of high thermal radiation output. The most likely reason for this is that the core of the galaxy is a dense star forming region. The spectral index becomes steeper moving further away from the centre of \NGC, indicating a dominance of non-thermal radiation which could be caused by either synchrotron or inverse-Compton radiative mechanisms. Objects which radiate using this mechanism include SNRs and energetic jets.

\FloatBarrier

\section{CONCLUSION}

We present and discuss a series of new highly sensitive, high resolution radio-continuum images of NGC~55 at wavelengths 20, 13, 6 and 3~cm. These images were created by combining data collected from observations using the ATCA radio interferometer telescope. As a result of combining the data, the final images had dramatically reduced levels of noise with higher angular resolution when compared with previous studies. From the new images, a total of 66 radio sources were detected within the field of NGC~55, 33 of which were detected over multiple wavelengths. Of these 66 sources, \NUMBKG{} have been classified as background sources, \NUMHII{} as \HII{} regions, and \NUMSNR{} as SNR candidates. Spectral indices were also calculated for these multi-wavelengh sources. A spectral index map was produced for the galaxy, revealing a high level of thermal radiation emission from the core of the galaxy. Several concentrations of high non-thermal radiation emission were also detected within the plane of the galaxy.

\section*{Acknowledgments}
The National Radio Astronomy Observatory is a facility of the National Science Foundation operated under cooperative agreement by Associated Universities, Inc. The Australia Telescope Compact Array is part of the Australia Telescope National Facility which is funded by the Commonwealth of Australia for operation as a National Facility managed by CSIRO. This paper includes archived data obtained through the Australia Telescope Online Archive (http://atoa.atnf.csiro.au) and the NRAO Science Data Archive (http://archive.nrao.edu).

\bibliographystyle{spr-mp-nameyear-cnd}
\bibliography{AOBrien_ngc55}
 
\end{document}